\documentstyle[prl,aps,twocolumn,epsf,psfig]{revtex}

\begin{document}

%\twocolumn[\hsize\textwidth\columnwidth\hsize\csname
%@twocolumnfalse\endcsname

\title{On the origin of magnetoresistance in Sr$_2$FeMoO$_6$}

\author{D.D. Sarma \cite{jnc} and Sugata Ray}

\address{Solid State and Structural Chemistry Unit, Indian Institute
of Science, Bangalore 560~012, INDIA \\}
%\date{today}

\author{K. Tanaka and A. Fujimori}

\address{Department of Physics and Department of Complexity Science and
Engineering, University of Tokyo, Bunkyo-ku, Tokyo 113-0033, Japan
\\}

\maketitle

%\newpage
\begin{abstract}
We report detailed magnetization ($M$) and magnetoresistance
($MR$) studies on a series of Sr$_2$FeMoO$_6$ samples with
independent control on anti-site defect and grain boundary
densities. These results, exhibiting a switching-like behavior of
$MR$ with $M$, establish that the $MR$ is controlled by the
magnetic polarization of grain boundary regions, rather than of
the grains within a resonant tunnelling mechanism.
\end{abstract}

PACS number(s): 75.47.Gk, 72.25.-b, 75.60.-d

%\vspace*{1cm}
\newpage

While colossal magnetoresistance (CMR) in manganites~\cite{book}
spurted a tremendous level of research activities in view of its
immense technological possibilities, the actual realization of any
device application has been mainly impeded by the requirement of a
low temperature and high magnetic field for an appreciable $MR$ to
be observed in these compounds. It has been
demonstrated~\cite{mathur,gupta} that substantial negative $MR$
can be achieved at reasonably low applied fields by specific grain
boundary engineering; however, the requirement of a low
temperature is still not conducive to wide-spread device
applications. Thus, a great deal of interest has naturally been
generated with the more recent discovery of CMR in polycrystalline
Sr$_2$FeMoO$_6$ at a considerably higher temperature and lower
magnetic field~\cite{koba}. Magnetic scattering at magnetic domain
boundaries can be ruled out as the origin of $MR$ in this
compound, because no significant $MR$ was observed in the single
crystalline compound~\cite{tokuraprb}, where magnetic domain
boundaries exist. While all observations clearly suggest
tunnelling magnetoresistance (TMR) to be the dominant cause of the
CMR in Sr$_2$FeMoO$_6$, there is no clear agreement or
understanding of the nature of tunnelling barriers in this system.
Since, a proper identification of the primary mechanism for $MR$
in this compound is essential as a first step towards realizing
any useful application, we have carried out specific experiments
to address this issue.

Two alternative origins of $MR$ in Sr$_2$FeMoO$_6$ have been
discussed so far. In one view, the physical grain boundaries are
believed to provide tunnel barriers. But, another source of such
tunnel barriers in this compound, arising from Fe/Mo antisite
disorder giving rise to antiferromagnetic, insulating Fe-O-Fe
patches in between fully-ordered metallic Sr$_2$FeMoO$_6$ islands
within a single grain, has also been convincingly put
forward~\cite{garcia}. Therefore, in this limit, the grain size of
the polycrystalline sample will not affect the TMR
significantly~\cite{garcia}. Various synthetic
parameters~\cite{font}, such as the annealing temperature, provide
a useful handle to change the Fe/Mo sublattice ordering, thereby
changing the density of antisite defects and the magnetization in
a systematic way. Unfortunately, these synthetic processes also
change the grain sizes (and thus the grain boundary density), the
average thickness of the grain boundaries and also possibly the
chemical composition of the grain boundaries (depending on the
oxygen partial pressure and temperature) in an uncontrolled
manner. Here, we propose a route that allows us to control the two
crucial parameters, namely the extent of ordering controlling the
antisite defect density and the grain size determining the grain
boundary density, independent of each other and keeping other
physical properties essentially unchanged, thereby providing a
critical test for the dominant mechanism of $MR$ in
Sr$_2$FeMoO$_6$. By varying these two control parameters
independently, we show that the origin and behavior of the
magnetoresistance is unique in this compound compared to all other
systems.

All the Sr$_2$FeMoO$_6$ samples, studied here were first prepared
by the arc melting method, leading to highly disordered samples,
as characterized by x-ray diffraction (XRD)~\cite{ssc}. However,
it produces samples (labelled A) with large grains (10-20~$\mu$m)
and relatively few grain boundaries, as confirmed by scanning
electron microscopy~(SEM). We also prepared three other samples,
B, C and D, by annealing A at 1173~K, 1523~K and 1673~K,
respectively, for a period of 5 hours. SEM in conjunction with
energy dispersive analysis of x-ray (EDX) revealed no change in
the grain morphologies and composition due to this annealing
process. The extent of the Fe/Mo ordering or the anti-site defect
density, quantified by the intensity of the supercell (111)
reflection at 19.6$^{\circ}$ normalized by the normal reflection
at 32.1$^{\circ}$
($I_{19.6^{\circ}}$/$I_{32.1^{\circ}}$)~\cite{koba,ssc}, increased
progressively for A, B, D, C.
($I_{19.6^{\circ}}$/$I_{32.1^{\circ}}$) for sample C turns out to
be larger than those previously prepared; consequently this sample
has one of the largest saturation magnetization
(3.54~$\mu_B$/f.u.) reported in bulk Sr$_2$FeMoO$_6$ samples so
far~\cite{garcia}. Then, we take approximately half of each of
these four samples and grind them to a fine particle size with an
average grain size of 2-3~$\mu$m and pelletize these at room
temperature to form samples E, F, G and H from samples A, B, C,
and D, respectively. Each pair of samples, namely (A, E), (B, F),
(C, G) and (D, H) thus have the same antisite defect density
within a group, but very different grain sizes and grain boundary
density.

In Fig. 1a we show the $M(H)$ plots (solid lines) for the large
grain samples, A-D, while for the small grain samples, we plot
-$M(H)$ in order to avoid crowding together of the data, measured
at 5 K. The saturation magnetizations of these samples increase
monotonically with the extent of Fe/Mo ordering, possibly
suggesting that antisite defects produce antiferromagnetic
regions~\cite{garcia}. Inset to Fig. 1b shows a typical change in
the resistivity ($\rho$) behavior  as a function of temperature
(T) within a pair of samples with the same antisite defect density
but different grain sizes. The resistivity of the large grain
sample is obviously metallic, being qualitatively and
quantitatively similar to that of the single crystal
samples~\cite{tokuraprb}. In contrast, the corresponding small
grain sample exhibits an insulating behavior with a resistivity
that is at least four orders of magnitude larger compared to the
large grain sample at all temperatures. Thus, it is clear that
grain boundaries offer insulating barrier to macroscopic charge
transport in these samples. Figure 1b shows the \%~$MR$
(=100$\times$[$R$($H$, T)-$R$(0, T)]/$R$(0, T)) for each of the
compounds at 20 K. For the purpose of clarity and easy comparison,
we show the \%~$MR (H)$ plot of the big grain samples A-D, for the
negative sweep of the magnetic field, while the \%~$MR (H)$ for
the small grain samples are shown for the positive sweep of the
applied field. The intrinsically more noisy nature of $MR$ in the
large grain samples arises from the very low resistance of these
samples (see inset) and a small $\Delta R$ = $R (H)$ - $R (0)$. It
is quite clear that the \%~$MR$ exhibited by the small grain
samples are uniformly several times larger than those of the
corresponding large grain samples. If we take into account the
fact that the resistivities of the small grain samples are at
least four orders of magnitude larger than that of the
corresponding large grain samples with identical anti-site defect
density, it becomes evident, that the absolute change in
resistivity, $\Delta R$, for the small grain sample is about five
orders of magnitude larger than that from large grain samples of
similar dimensions on application of the same magnetic field.
These facts clearly establish that the predominant contribution to
the observed $MR$ in these samples arises from grain boundary
effects, rather than from antisite defects.

In order to quantify the comparisons, we plot the \%~$MR$ at 20 K
and 1 T as a function of the ordering in Fig. 2. Since the extent
of ordering is a direct measure of the antisite defect density,
Fig. 2, evidently showing a very large enhancement (typically
100-200 \%) of $MR$ for the small grain samples compared to the
large grain ones at about the same level of ordering, clearly
establishes the grain boundaries instead of the antisite defects
as the primary source for the dominant mechanism of $MR$ in
Sr$_2$FeMoO$_6$. Interestingly, there is a small increase in $MR$
of sample D compared to sample C, though the extent of ordering as
well as the saturation magnetization (see Fig. 1a) are lower for
the sample D. In these large grain, molten ingots, the effect of
grain boundaries is expected to be minimized, thereby enhancing
the chances of observing small contributions from other effects,
such as that of antisite defects. It is indeed true that the
antisite defect density is much larger in sample D compared to C,
owing to a lower Fe/Mo ordering; therefore, the small increase in
$MR$ observed between samples C and D suggests a small, but finite
contribution to $MR$ from such antisite defects, while the
predominant part of the $MR$ in all samples arises from the grain
boundary effects. A similar argument would suggest a higher $MR$
for sample B compared to that of D and the highest $MR$ in A among
the large grain samples due to a systematic increase of antisite
defects, in sharp contrast to the experimental results (Fig. 2).
This suggests that excessive introduction of antisite defects has
an adverse effect on the $MR$, most likely arising from a marked
reduction of the magnetic polarization of the bulk and hence the
spin polarization at $E_F$~\cite{tanu}.

In order to understand the nature of $MR$ in these compounds, we
plot the \%~$MR$ as a function of $M$/$M_s$ in Fig. 3; the results
for the corresponding pairs (A, E), (B, F), (C, G) and (D, H) are
shown in different panels. These plots make a number of
interesting behavior obvious. Evidently, the \%~$MR$ does not
exhibit the expected $(M/M_s)^2$~\cite{pinaki} behavior of a
tunnelling barrier, except for the lowest $MR$ (\%~$MR$~$\le$~1\%)
regime in the large grain sample, A-D, as shown more clearly in
the insets with \%~$MR$ plotted against $(M/M_s)^2$, also
illustrated by overlapping the best fit $(M/M_s)^2$ function
(dotted line) on the experimental plots in the main frames of Fig.
3. Manifestation of the $(M/M_s)^2$ behavior for the initial
($\le$ 1\%) $MR$ specifically in the large grain samples, with
significantly reduced grain boundary effects, suggests that the
tunnelling barriers are possibly provided for these samples by the
previously discussed antisite defect boundaries~\cite{garcia}. The
remaining and the larger part of the $MR$, therefore, arises from
intergrain effects. However, this intergrain tunnelling
contribution evidently does not have the expected $(M/M_s)^2$
dependence and intriguingly almost switches on (signalled by the
very rapid increase) at a substantially large value of $(M/M_s)$
in every case, though more pronounced in the highly ordered small
grain samples. For example, sample G with the maximum $MR$ value
(see Fig. 3c) exhibits only about 6~\% $MR$ for $(M/M_s)$~=~0.9,
the remaining nearly 18~\% $MR$ is achieved with the magnetization
achieving the last 10~\% of its saturation value. Though much
attention has been focussed on the {\emph low-field} $MR$ in
Sr$_2$FeMoO$_6$, in view of the results in Fig. 3 and the above
discussion, it appears that Sr$_2$FeMoO$_6$ in fact requires an
unusually large field to realize the $MR$ in the sense that an
applied field, though enough to reach 50-80~\% of the saturation
magnetization, is not enough to show more than 1~\% $MR$. These
observations clearly suggest that the dominant origin of $MR$ in
Sr$_2$FeMoO$_6$ cannot be understood in terms of the usual TMR
scenario with electron transport across an inert tunnelling
barrier by aligning the magnetization of the two grains across the
grain boundary by the application of a magnetic field. It appears
to be necessary to magnetize and therefore magnetically align the
grain boundary itself in order to achieve the $MR$ response,
suggesting the switching-on of a resonant tunnelling mechanism in
the grain boundary magnetization, rendering the \%~$MR$ relatively
insensitive to the extent of bulk magnetization, $(M/M_s)$. From
the low-field value of ($M/M_s$) $\sim$ 0.9, it seems that roughly
about 10 \% of the total magnetization in the high field comes
from the magnetic nature of the grain boundary. A final evidence
for these suggestions comes from the following observation.

The $MR$ plots for the small grain samples consistently exhibit a
sizable hysteresis in the low field ($\le$ 1 T) regime, while that
of the large grain samples do not exhibit any noticeable
hysteresis (see Fig. 1b and 3). The near absence of hysteresis in
the large grain samples is consistent with the near absence of any
hysteresis in the corresponding magnetization plots in Fig. 1a. It
has been found for all oxide systems, such as for CrO$_2$ in ref.
[12], for manganites in ref. [13], as well as for Co-Cu alloys in
ref. [14] that the peak in $MR$ coincides with the corresponding
coercive fields ($H_c$), as should indeed be expected for TMR. In
order to carefully compare the hysteresis in $MR$ and $M$ in small
grain samples, we show $MR$ and $M$ as function of $H$ on an
expanded scale ($|H|\le$~0.3~T), in Fig. 4. The figure clearly
shows the hysteresis loops in $M$ vs. $H$ as well as in $MR$ vs.
$H$. In striking contrast to other systems~\cite{coey,cnr,xiao},
small grain Sr$_2$FeMoO$_6$ intriguingly shows a peak in $MR$ at
an $H$ about 6 times larger than $H_c$. This is possible only if
the grain boundary region controlling the TMR in this material has
a much larger coercive field than $H_c$ of the bulk, while the
intra-grain properties dominate the $M(H)$ behavior. This
suggestion is consistent with all the observations presented here
so far. Additionally this is further supported by a small but
definite increase in the coercive fields of the small grain
samples compared to the large grain ones, due to the presence of
large number of grain boundaries with much higher $H_c$ in the
small grain samples, as illustrated for the (D, H) pair in the
inset of Fig. 4.

In conclusion, by controlling independently the anti-site defect
density and grain boundary density in a series of Sr$_2$FeMoO$_6$
samples, we show that the $MR$ originates primarily from
tunnelling across the grain boundaries, with very small
contributions from anti-site defects. This TMR from the grain
boundaries exhibits several intriguing features, most notably (i)
a switching-like behavior of $MR$, and (ii) the peak in $MR(H)$
appearing at an $H$ several times higher than the $H_c$ seen in
$M(H)$ curves. These are explained in terms of the different,
specifically harder, magnetic nature of the grain boundary region
compared to the bulk, controlling the resonant tunnelling behavior
and thereby the $MR$, instead of providing a simple tunnelling
barrier.

We thank A. Fujiwara for help in the SQUID measurements. DDS
thanks university of Tokyo for hospitality during a part of this
work.

%\newpage

%\begin{table}

%Table~I

%{\bf Big grain samples}

%\begin{tabular}{|c| |c| |c| |c| |c|}
%Sample & Annealing Temperature ($^{\circ}$ C)& \%Order & $M_s$ at
%5 K and 5.5 T& \%$MR$ at 20 K and 7 T
%\\ \hline A & No annealing & 20.3 & 1.04 & -7.16
%\\ B & 900 & 22.1 & 1.42 & -9.73
%\\ C & 1250 & 93.8 & 3.54 & -9.44
%\\ D & 1400 & 58.8 & 3.12 & -7.25
%\\ Sr$_2$FeMo$_{0.3}$W$_{0.7}$O$_6$  & 2.15 & -7.3$\times$ 10$^{-2}$ & 2.07

%\end{tabular}
%\end{table}

%\begin{table}

%{\bf Small grain samples}

%\begin{tabular}{|c| |c| |c| |c| |c|}
%Sample & Annealing Temperature ($^{\circ}$ C) & \%Order & $M_s$ at
%5 K and 5.5 T & \%$MR$ at 20 K and 7 T
%\\ \hline E & No annealing & 13.9 & 1.19 & -12.31
%\\ F & 900 & 22.4 & 1.36 & -15.74
%\\ G & 1250 & 93.5 & 3.28 & -24.43
%\\ H & 1400 & 57.5 & 2.48 & -21.07

%\end{tabular}
%\end{table}

\begin{center}
\begin{figure}
\caption{(a) $M(H)$ plots (solid lines) and -$M(H)$ plots (dashed
lines) for the large and small grain samples, respectively. (b)
\%~$MR(H)$ are shown for the large grain samples (solid circles)
on the left half and for the small grain samples (lines) on the
right half. The inset shows the typical changes in the resistivity
with temperature between a large and a small grain samples with
the same level of anti-site defect densities.}
\end{figure}
\end{center}

\begin{center}
\begin{figure}
\caption{Variation of \%~$MR$ at 1T with the extent of ordering in
the sample, as measured by the normalized intensity of the
supercell reflection at 19.6$^\circ$.}
\end{figure}
\end{center}

\begin{center}
\begin{figure}
\caption{The dependence of \%~$MR$ on $M/M_s$ for various samples.
Each of the four panels compare the \%~$MR$ obtained for a pair of
large and small grain samples with the same anti-site defect
density. The insets show the \%~$MR$ as a function of the
$(M/M_s)^2$ for the corresponding large grain samples, with the
best-fits to ($M/M_s)^2$ dependence, expected theoretically, shown
by the solid lines in the inset and the dashed lines in the main
frame.}
\end{figure}
\end{center}

\begin{center}
\begin{figure}
\caption{A comparison between the low-field region of the $MR$ and
$M$ in one illustrative case. The inset shows the comparison of
$M(H)$ for a typical large and small grain samples.}
\end{figure}
\end{center}

%\pagebreak
%\begin{figure}[h]
%\vspace*{-1.3 in}
%\centerline{\epsfysize=6.0in\epsffile{fig1.eps}}
%\vspace*{-3.0 in}
%\end{figure}

%\pagebreak
%\begin{figure}[h]
%\vspace*{1.5 in} \centerline{\epsfysize=5.5in\epsffile{fig2.eps}}
%\vspace*{0.0 in}
%\end{figure}

%\pagebreak
%\begin{figure}[h]
%\vspace*{1.5 in} \centerline{\epsfysize=5.5in\epsffile{fig3.eps}}
%\vspace*{0.0 in}
%\end{figure}

%\pagebreak
%\begin{figure}[h]
%\vspace*{1.9 in} \centerline{\epsfysize=5.0in\epsffile{fig4.eps}}
%\vspace*{0.0 in}
%\end{figure}

\end{document}